# Stationary and Dynamic Reference Frame Comparison Based Microgrid Application

*ILYAS BENNIA [1,2*] , YACINE DAILI [1,2] , ABDELGHANI HARRAG [1,2] , WALID ISSA[3]*

[1]Renewable Energy Deployment and Integration Team, Mechatronics Laboratory (LMETR), Optics and Precision Mechanics Institute, Ferhat Abbas University Setif 1, Setif, Algeria

[2]Electrotechnics Department, Faculty of Technology, Ferhat Abbas University Setif 1, Setif, Algeria

[3]Engineering & Maths Department, Sheffield Hallam University, Sheffield S1 1WB, UK

*Abstract*—**This paper presents a brief comparison for voltage and current controllers implementation in both stationary and dynamic reference frame for a microgrid (MG) application. Diagrams of implementations are reviewed and the simulation results are presented to show the performance of each topology.**

## II. INTRODUCTION

Non-renewable resources, such as diesel, coal, and gas, are the primary energy sources of electrical energy produced by traditional power generators worldwide. However, the increasing demand for electricity, consumption of reserves of non-renewable resources, and generation of electrical energy from non-renewable resources conducted to environmental pollution. Therefore, the development of a distributed generation (DG) system using renewable resources such as solar and wind energy to produce electricity is necessary.

These renewable resources are difficult to integrate and exploit due to the stochastic behavior of their prime movers. A microgrid is an optimal solution to integrate and control distributed renewable resources and tied them to the utility grid. Using low-voltage distribution system consisting of DG units, energy storage devices, and load [1]. Additionally MG is able to operate in island or connected to a main distribution system, fig. 1 shows the general architecture of a MG[2]. In addition, compared with a single DG unit, a MG has high capacity and control flexibility to fulfill power-quality requirements [3].

Hierarchical control of MG was introduced in to deal with control challenges of MG and to endow smartness and flexibility to MGs [4], The hierarchical control based on three layers The primary layer include droop method [3] and virtual impedance loop [5], which are responsible for providing the voltage reference for inner loops which are often considered as level zero, inner loops are responsible for voltage and current regulating by a direct interaction with the VSIs, the secondary layer [6] allows the restoration of the deviations produced by the primary control and the tertiary layer manages the power flow between the MG and electrical distribution system [7].

The inner loops are responsible for voltage and current regulating by a direct interaction with the VSIs which are the bloc stones and the significant components of a microgrid, inner loops consist of a cascaded control structure using a voltage and current controllers followed by a pulse-width-modulator (PWM). Here, the control loops regulate inverter state variables by computing an output voltage reference, which is the modulated signal by the PWM technic[8].

Inner loops can be implemented in a dq reference frame as described in [9]. using a conventional PI controller for a microgrid application although it can be implemented in αβ stationary reference frame as studied in [10][11] using PR resonant controller which is adopted in many literature.

This paper is organized as follow. Primary control is illustrated in section two, Inner loops were reviewed in section three for either stationary and dynamic reference frame with a brief comparison to each other, simulation results in section four and the paper is concluded by section five.

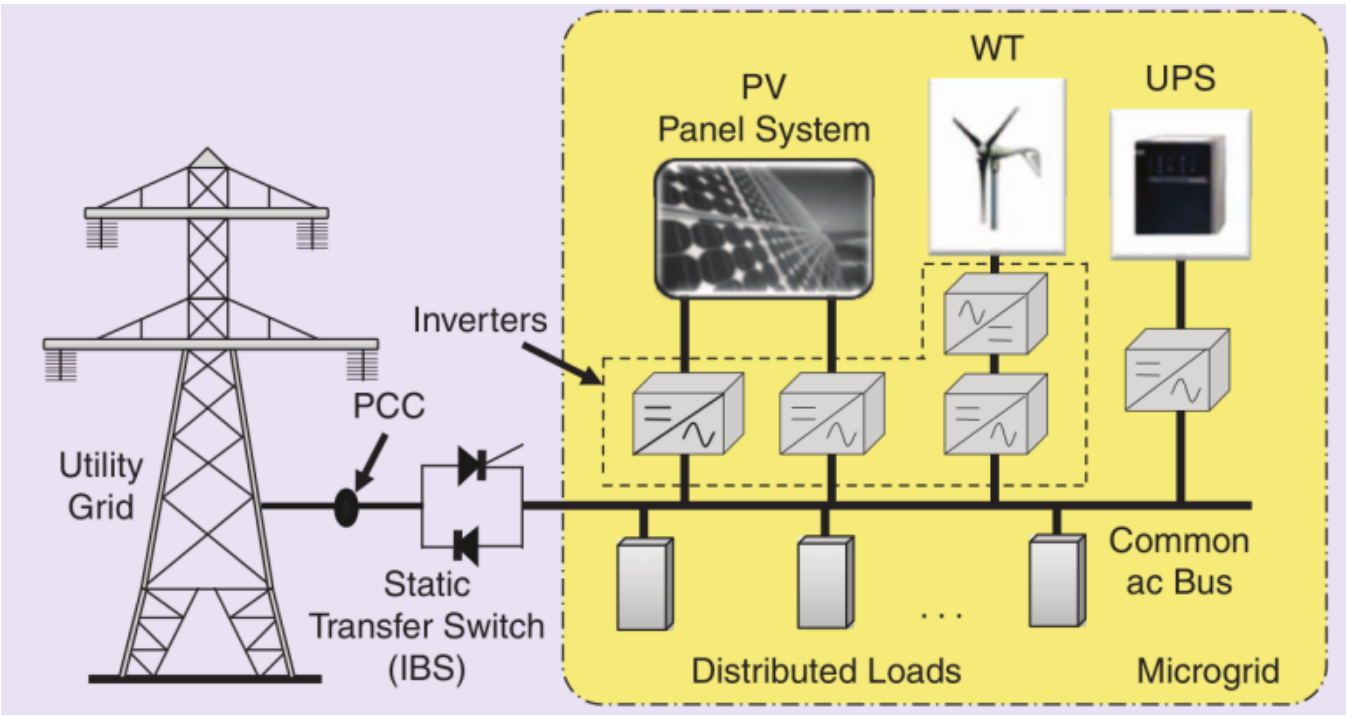

Figure 1. Gneral architecture of a microgrid

## III. PRIMARY CONTROL

Implemented locally primary control is responsible for power sharing, voltage and frequency regulation either in grid forming or grid following mode. This hierarchical layer sends control in intervals of several milliseconds to the VSIs to maintain the MG stability, communication in this control layer is undesired [12], the primary control is divided in two: inner loops responsible for regulating the power converters' voltage and current output, and the outer loop in charge of ensuring safe and correct power-sharing, inner loops have been widely studied in the literature however this topic still subject to research for enhancing the robustness against uncertainties[13], improving dynamic response, reducing unbalances and harmonic mitigation[14], developing control schemes plans capable of operating in both grid and islanded modes and providing a smooth transition for MG operation modes. Outer loops are represented by the well-known droop control technique [15] and virtual impedance loop, droop control has the ability to adjust the frequency and the voltage amplitude of the voltage reference according to the active and reactive powers ensuring P and Q flow which is lead for power-sharing between VSIs, virtual impedance loop is added to improve the current sharing by fixing and normalizing the output impedance, figure (2) shows the primary control more details are mentioned in [16].

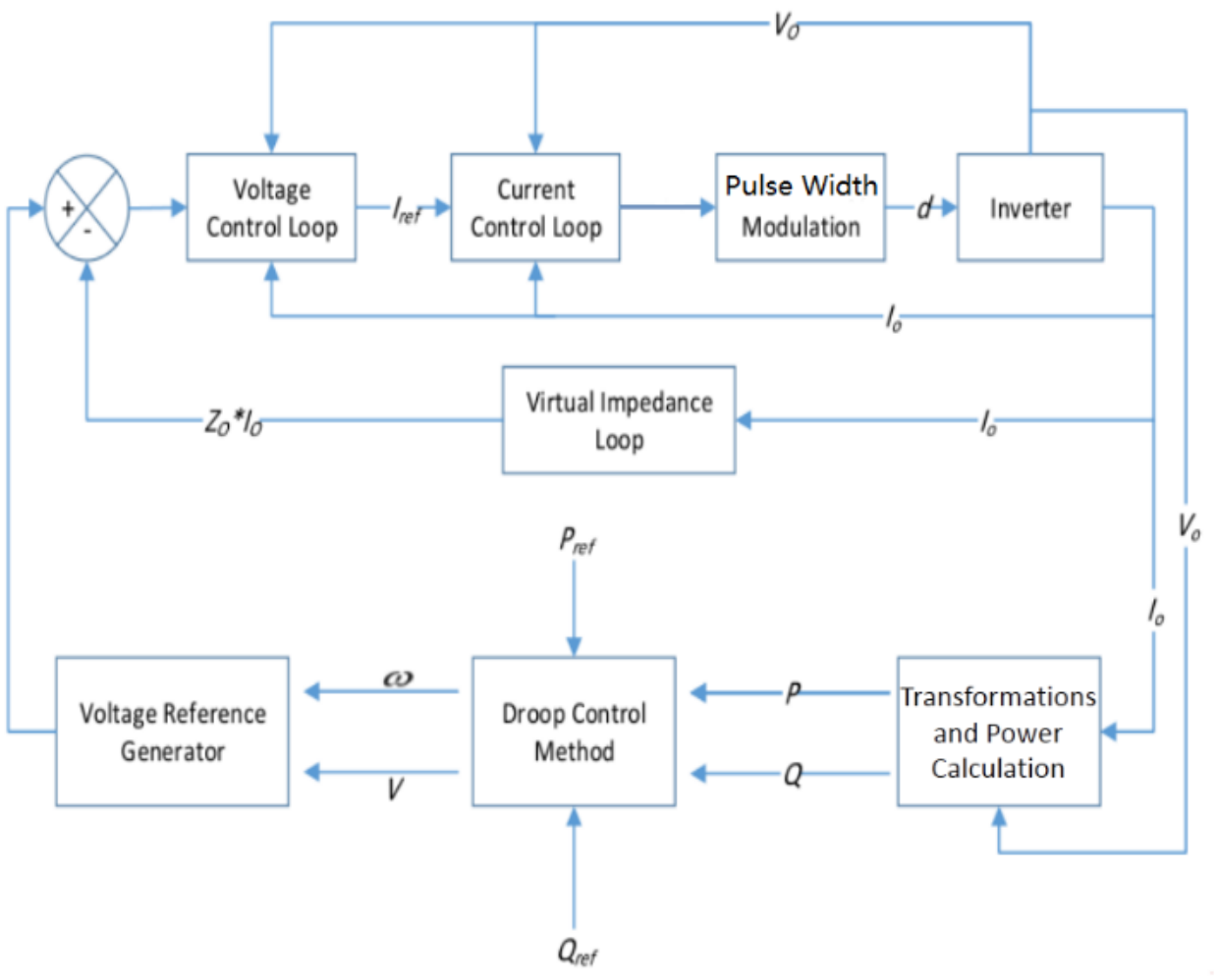

Figure 2. Primary control structure

## IV. INNER LOOPS

The inner control loops are known as level -zero voltage and current loops, it consists of two controllers for ensuring the MG stability, the voltage controller received the voltage reference generated by the droop control and virtual impedance loop, which in turn generate the current reference for the current controller, this latter provides the voltage reference for the pulse width modulation bloc, In this section, inner loops controllers are reviewed in terms of the reference frame they are implemented[10].

### A. *Based αβ on stationary reference frame*

Stationary reference frame control, based on the abc/αβtransformation, a three-phase system can be represented in two independent single-phase systems. Generally the voltage and current controllers are based on a Proportional-resonant (PR) controller, where generalized integrators (GI) are used to perform zero steady-state error, transfer functions of voltage and current controllers expressed follow:

$$G_v\ (s) = K_{pv} + \frac{K_{iv}s}{s^2+w_0} \tag{1}$$

$$G_i\ (s) = K_{pi} + \frac{K_{ii}s}{s^2+w_0} \tag{2}$$

Where Kpv and Kpi are the proportional gains and Kiv, Kii are the integral gains, and $w_0$ denotes resonant frequency. The PR controller shows a high gain around the resonant frequency, which is tuned at the fundamental MG frequency[17]. As PR controller depends on the frequency, it should be carefully designed considering frequency response because even a small deviation of system frequency can reduce the advantage of its high gain around the resonant frequency. Thus, a frequency adaptive PR controller is proposed in [18], whose resonant frequency is estimated by a PLL instead of being fixed. In addition, harmonic compensator given in [10]by:

$$G_h\ (s) = \sum_h \frac{K_{ih}s}{s^2+(hw_0)^2} \tag{3}$$

Can be added in parallel with the PR controller for the fundamental grid frequency, where h is the order of the harmonic to be attenuated. This helps minimize the corresponding harmonic components in this paper harmonic compensators are not included. Figure (3) shows the block diagram of the inner control loops of a three-phase VSI in αβ reference frame.

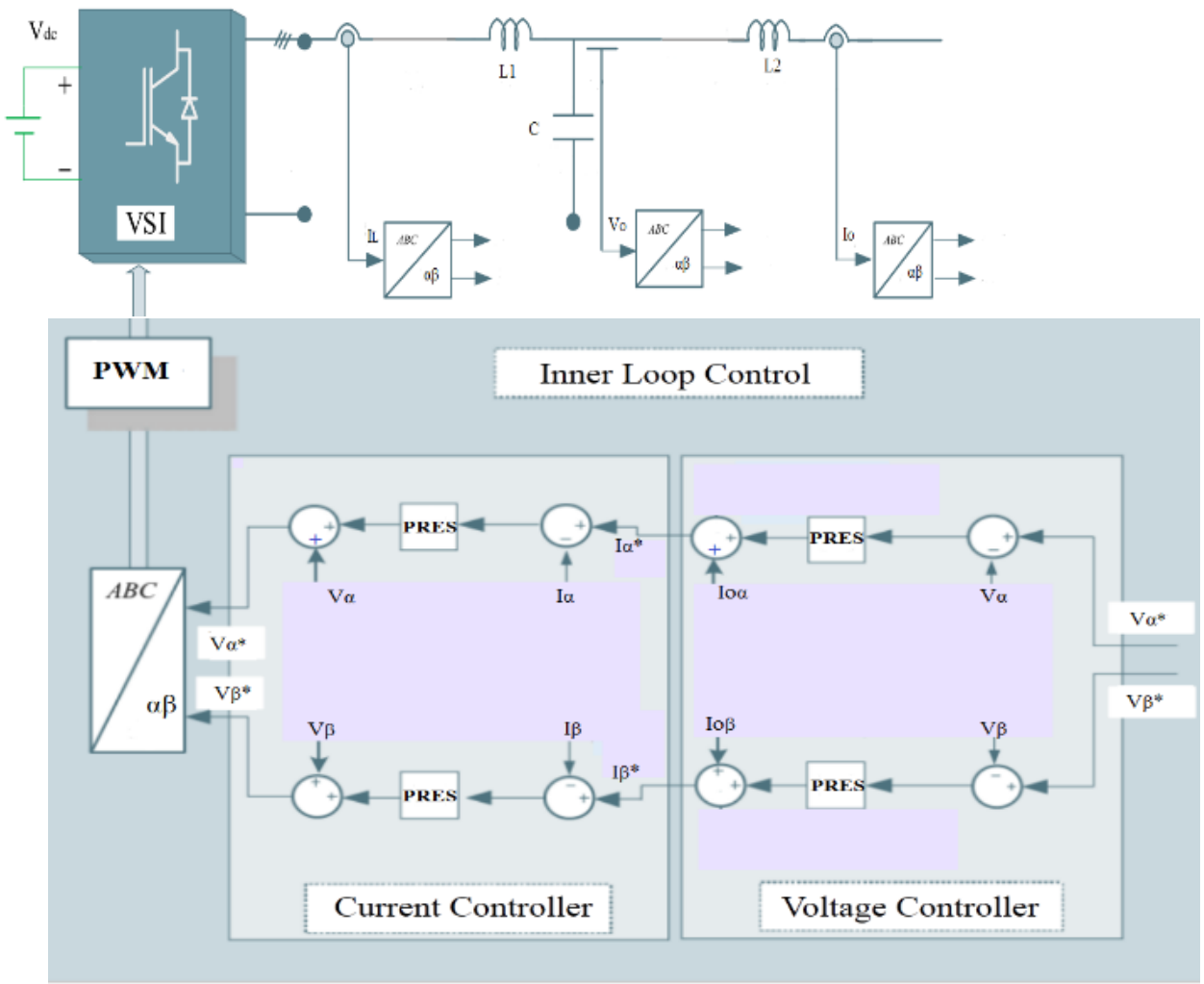

Figure 3. Voltage and current controllers in αβ refrence frame

Notice that the power calculation in αβ frame using following equations:

$$p = v\alpha \cdot i\,o\alpha + v\beta \cdot i\,o\beta \tag{4}$$

$$q = v\beta \cdot i\,o\alpha - v\alpha \cdot i\,o\beta \tag{5}$$

### B. *Based on dqdynamicreference frame*

Synchronous or dq reference frame, it converts grid voltage and current into a frame that rotates synchronously with the MG voltage vector by Park Transformation, three-phase time-varying signals are transformed into DC signals. In dq-control, PI controllers are typically used considering their capability of regulating direct signals without steady-state error[17]. The transfer’s functions of a PI controllers are described follow:

$$G_{PIvoltage}\ (s) = K_{pv} + \frac{K_{iv}}{s} \quad (6)$$

$$G_{PIcurrent}\ (s) = K_{pi} + \frac{K_{ii}}{s} \quad (7)$$

Where Kpv and Kpi are referred to proportional gains and Kvi, Kii are the integral gains, respectively. Established on symmetrical component theory, a voltage vector can be divided into positive-sequence component rotating anti-clockwise and negative-sequence component rotating clockwise. If a set of unbalanced signals is transformed into positive-sequence reference frame, the negative sequence component will be noted as AC signals oscillating at double the fundamental frequency same thing if positive-sequence components are transformed into negative-sequence reference frame. Because of this coupling between the two synchronous reference frames, PI controllers remain helpless to give satisfactory performance under unbalanced conditions[17]. This method requires two of PI regulators, one regulating only positive- sequence current in positive-sequence reference frame and the second controlling only negative-sequence current in negative-sequence reference frame, which allows the current in different sequences to be controlled within its own reference frame, notice that the capacity of harmonic compensation using PI controllers, is very low figure (4) shows the implementation of inner control loops in dq reference frame using PI controllers where Voltage feed forward and cross coupling is required , notice that the power calculation in dq frame as shown in the following equations:

$$p = vd \cdot i\,od + v\,q \cdot ioq \quad (8)$$

$$q = vd \cdot i\,oq - vq \cdot i\,od \quad (9)$$

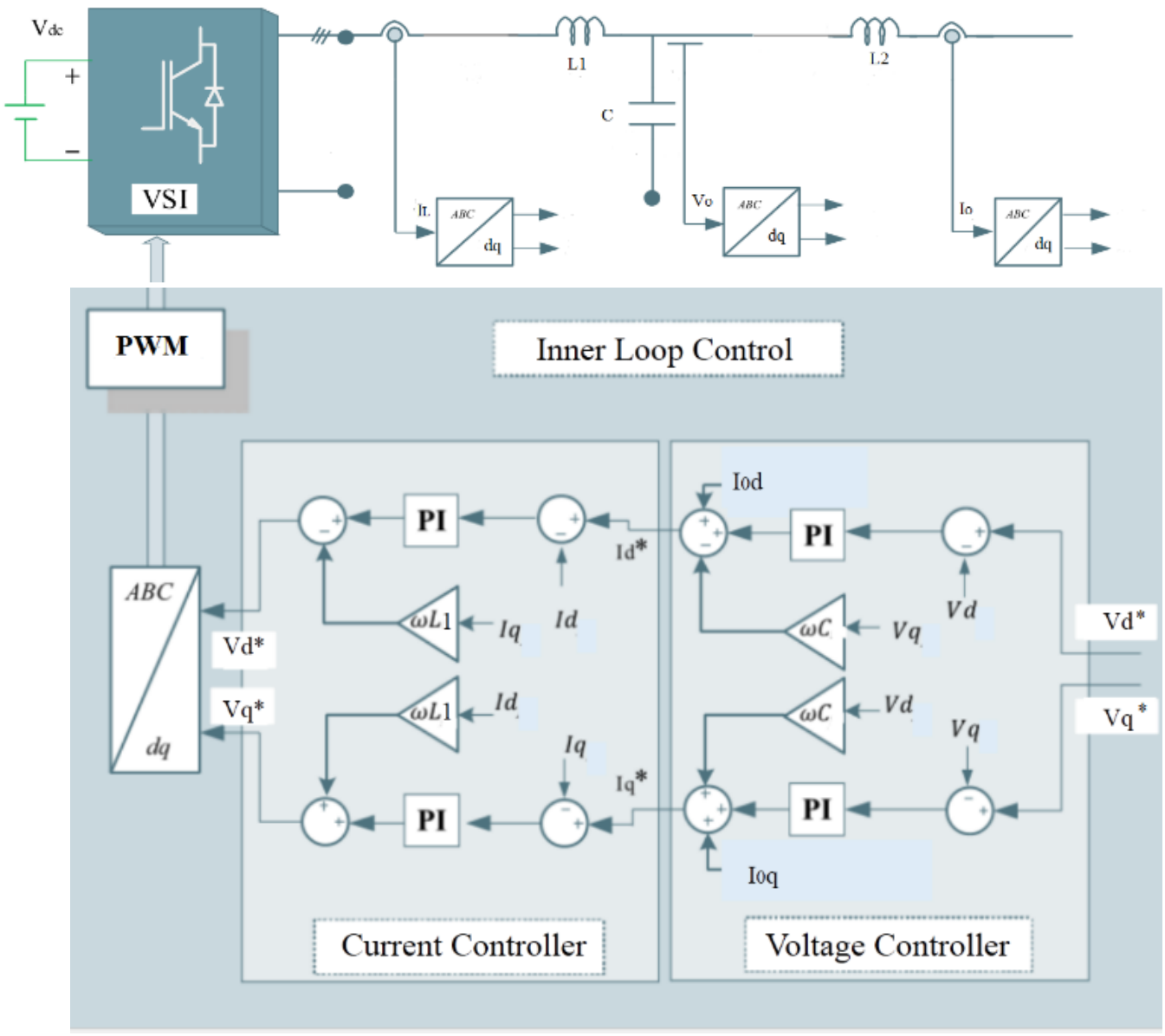

Figure 4. Voltage and current controllers in dq refrence frame

## V. SIMULATION RESULTS

In order to compare the performances of the voltage and current controllers under αβ reference frame and dq reference frame a simulation was done using the control loops depicted in figure (3) for the αβ reference frame and figure (4) for the dq reference frame both of them attached to a resistive load by MTLAB-Simulink software without including virtual impedance, the parameters are described in TABLE I.

Figure (5) and (10) show the tracking capibilities for both voltage controllers PR resonators in αβ frame and PI in dq frame where they achieve a zero steady-state error with a negligeable overshoot for the topology using PR resonant controller, same thing for current controllers in figure (6) and (11) where there is a high match between the refrence current and the output current, voltage outputs are shown in figure (7) and figure (12) they present a fast dynamic and smooth response and reached the steady-state which confirme the effictivness of the both regulators in the tow references frames.The measured active power in stationary reference frame reached the nominal power which is more accurate than the measured one in the dynamic reference frame notice that the technic of measuermant is different in each topology as afromentionned.Total harmonic destoration presnt in the stationary frame usin PR regulator 1.19% is bigger than the amount of it in the dynamic refrence using PI controllers 0.41%.

TABLE I. MICROGRID AND CONTROLLERS PARAMETERS

| Parameter name | Acronym | Value | Units |
|---|---|---|---|
| Nominal AC voltage | V | 220 rms | volt |
| Nominal frequency | f | 50 | Hertz |
| Load resistance | R | 145.2 | ohm |
| Filter output inductance | L1 | 0.003 | H |
| Filter capacitance | C | 4.8162e-06 | F |
| Filter inductance | L2 | 0.001 | H |
| **Droop control parameters** | | | |
| Proportional frequency droop | Kw | 0.0003 | W/rd |
| Proportional amplitude droop | Kv | 0.004 | Var/V |
| **αβ frame** | | | |
| Proportional gain PRES voltage compensator | Kp v | 0.2 | A/V |
| Integralgain PRESvoltage compensator | Ki v | 100 | A/(Vs) |
| Proportional PRES gain current compensator | Kp i | 5 | $A^{-1}$ |
| Integral gain PRES current compensator | Ki i | 400 | $(As)^{-1}$ |
| Cut-off frequency | $W_0$ | 0.03 × ω nom | rad/s |
| **dq frame** | | | |
| Proportional gainvoltage compensator | Kp v | 0.1 | A/V |
| Integral gain voltage compensator | Ki v | 3 | A/(Vs) |
| Proportional gain current compensator | Kp i | 20 | $A^{-1}$ |
| Integral gain current compensator | Ki i | 1000 | $(As)^{-1}$ |

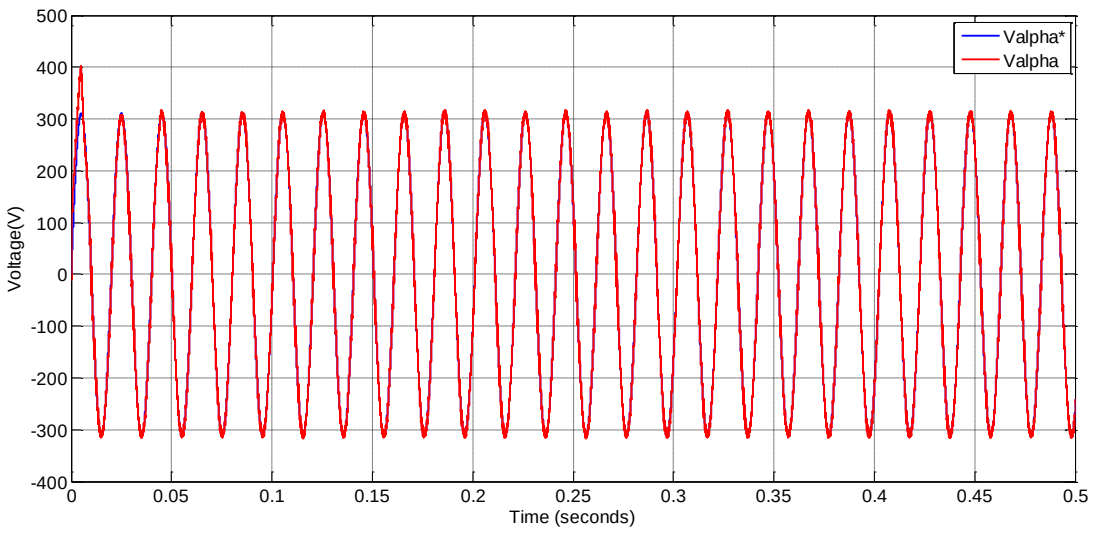

Figure 5. Valpha voltage tracking from voltage control loop

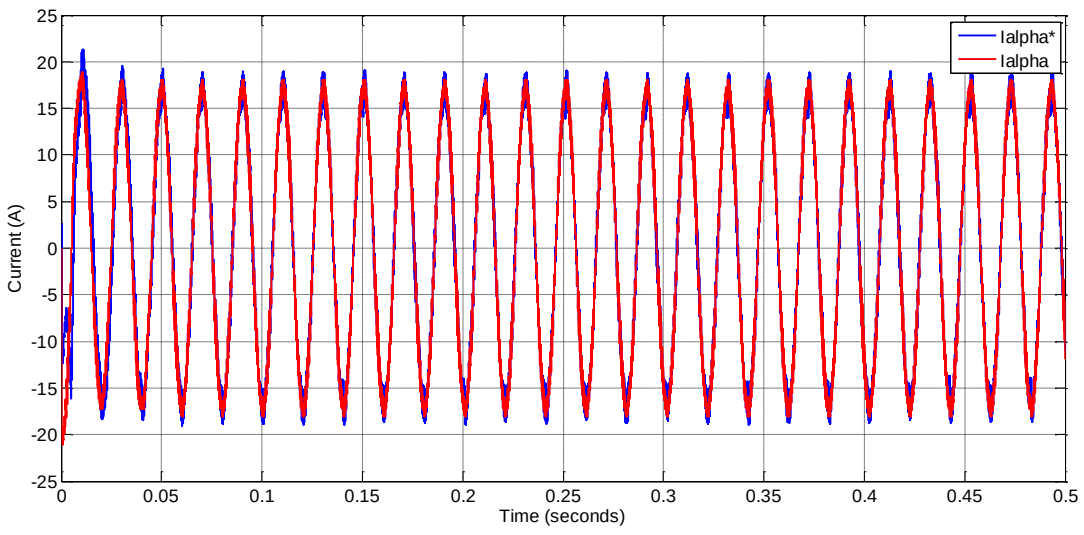

Figure 6. Ialpha current tracking from current control loop

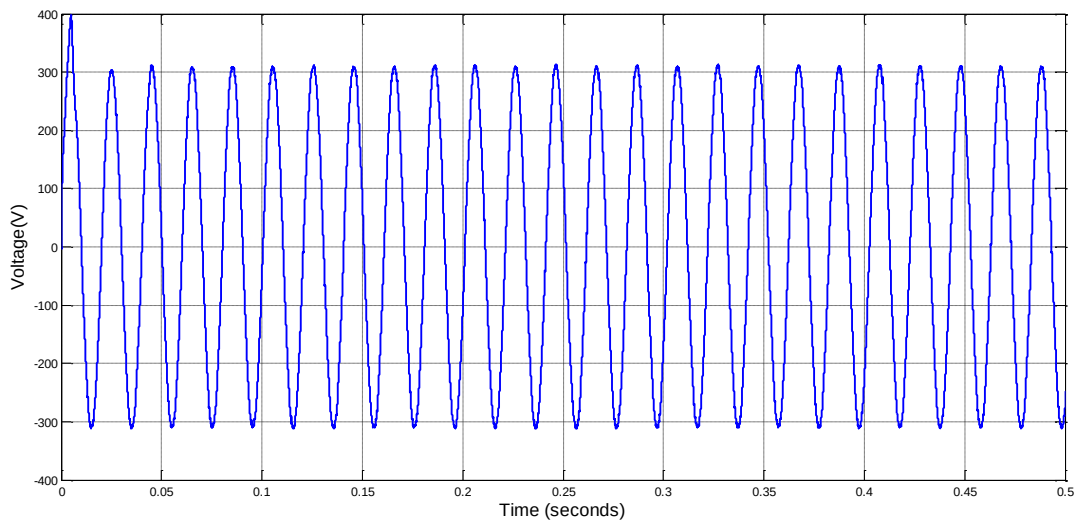

Figure 7. Output voltage phase A (αβ framework)

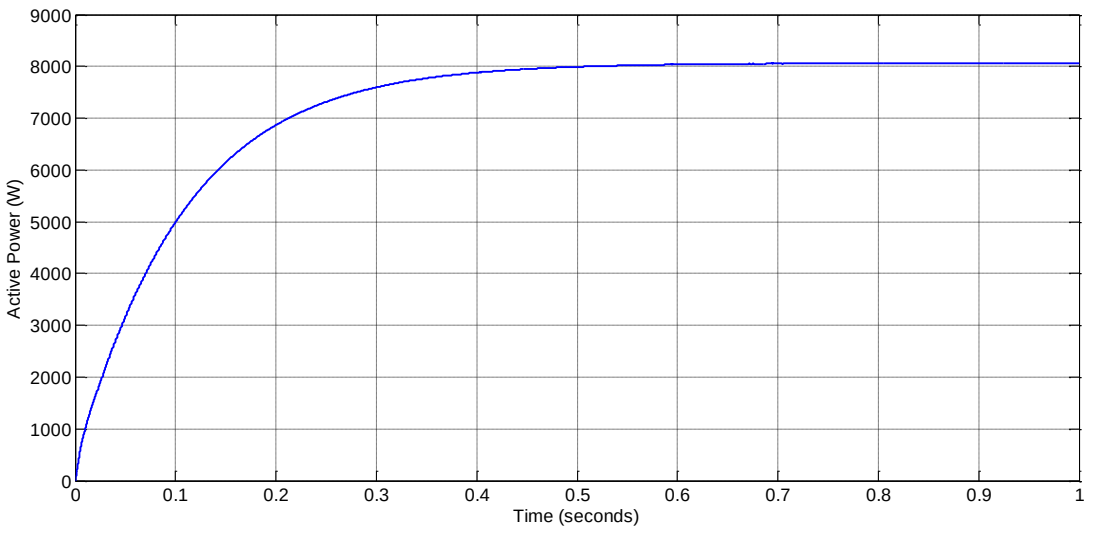

Figure 8. Active power curve (αβ framework)

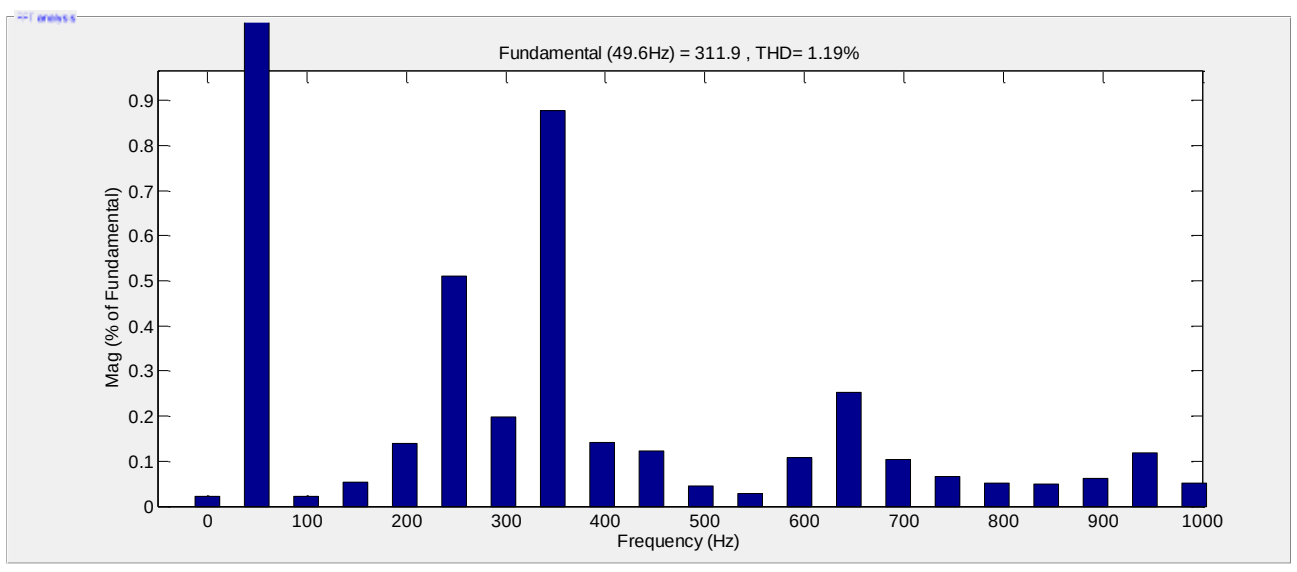

Figure 9. Spectrum frequency of output voltage in αβ framework

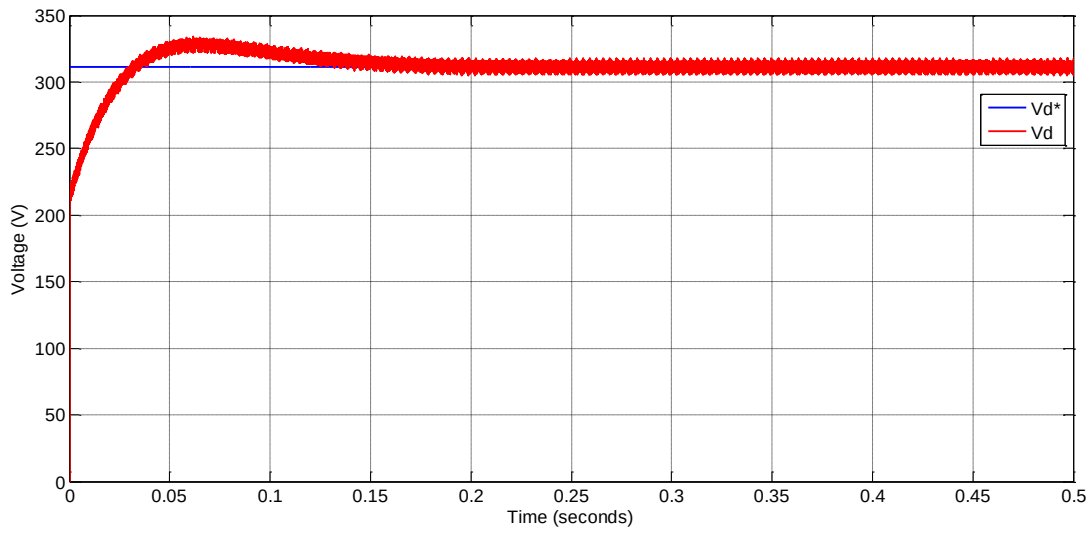

Figure 10. Vd voltage tracking from voltage control loop

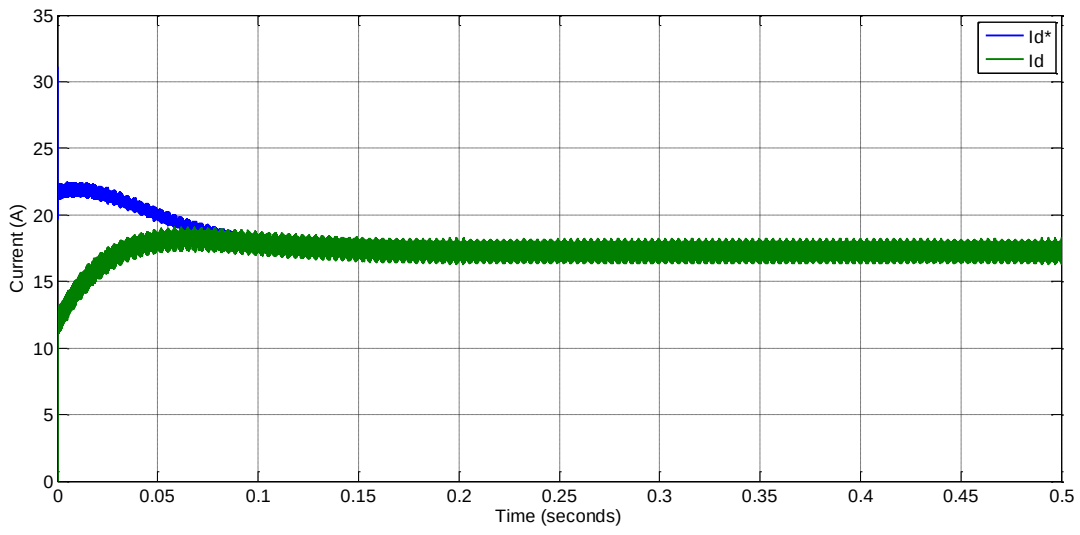

Figure 11. Id current tracking from current control loop

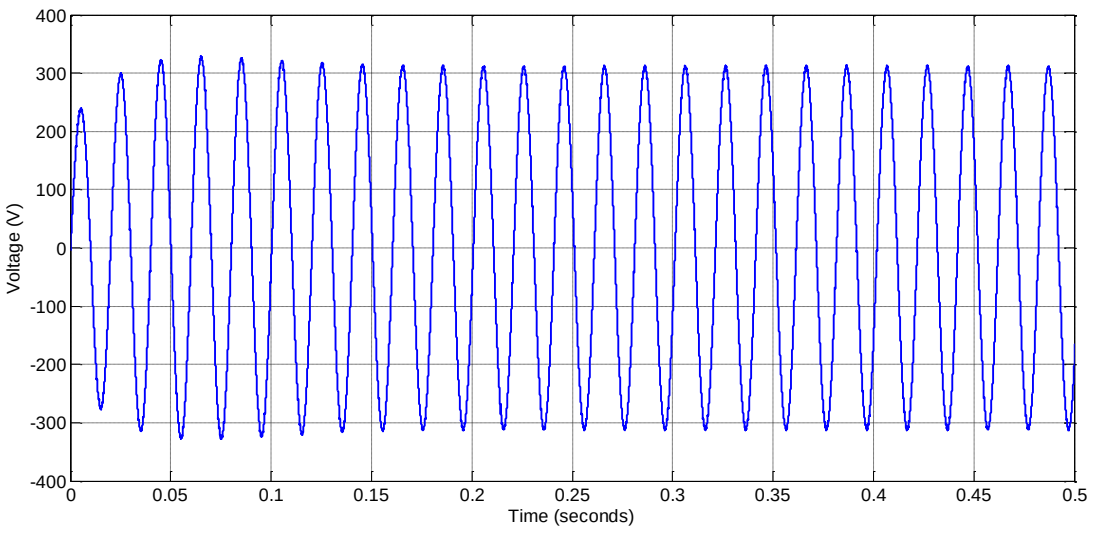

Figure 12. Output voltage phase A (dq framework)

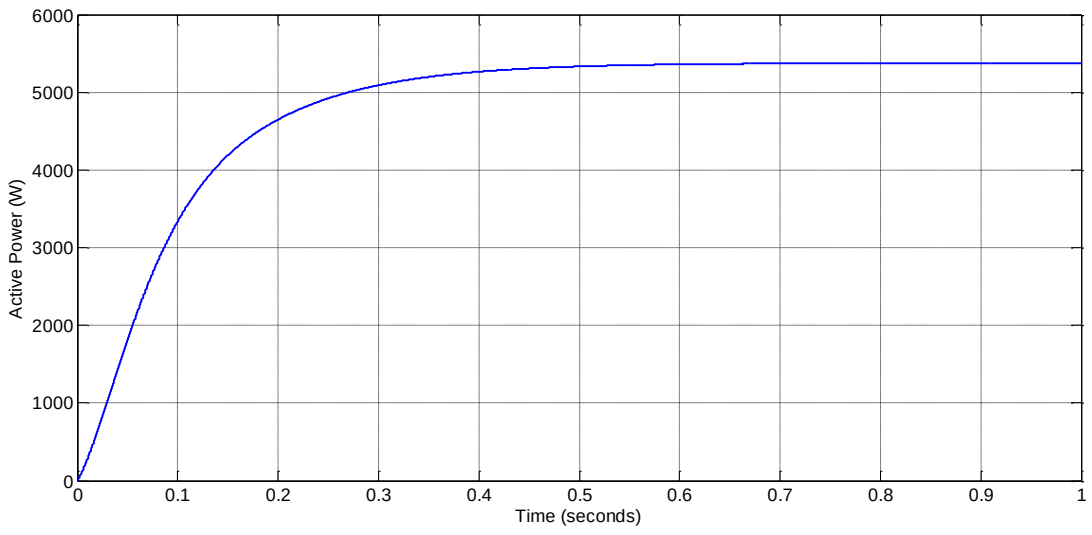

Figure 13. Active power curve (dq framework)

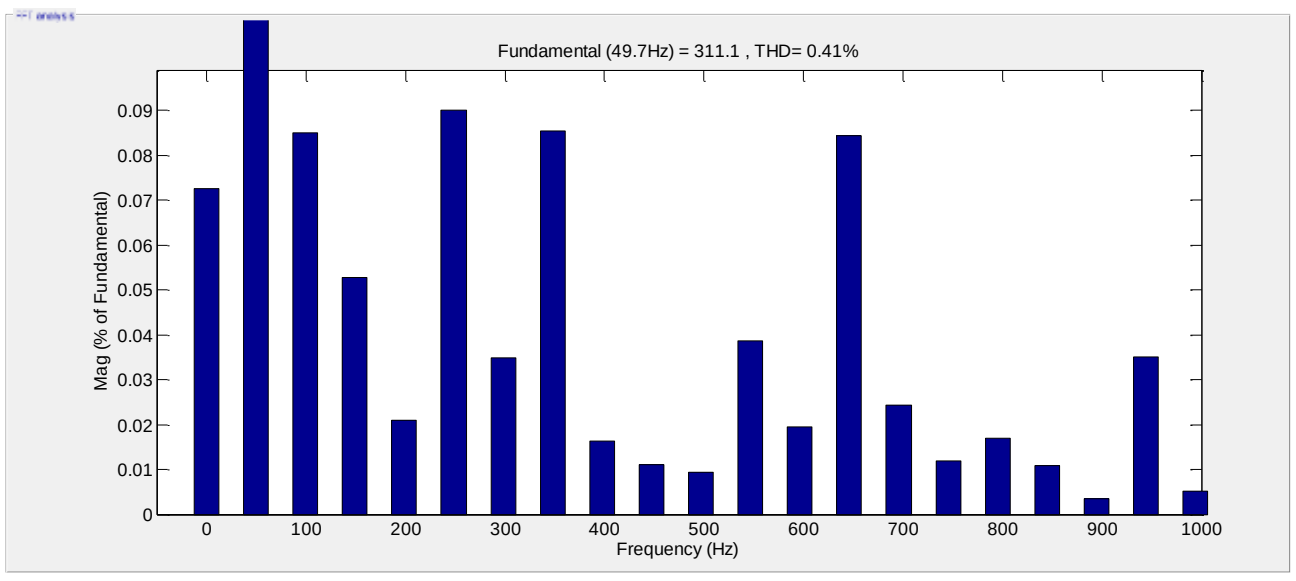

Figure 14. Spectrum frequency of output voltage in dq framework

## VI. DESCUSSION AND CONCLUSION

In this paper we presented a brief comparison between the stationary and the dynamic reference frame for MG application, the primary control is described and the implementation of inner loops in both αβ and dq frame is reviewed, as explained the structure implementation is simple with using of PR controllers when comparing the use of synchronous frame. There is no error in steady state condition when using either PR controllers or PI controllers, THD is more present when using PR controller, TABLE II shows a general comparison adopting many features.

TABLE II. GENERAL COMPARAISON OF USING STATIONARY AND DYNAMIC FRAMEWORK IN MICROGRIDS IMPLEMENTATION

| Frames | αβ frame | dq frame |
|---|---|---|
| Accuracy | Strong | High |
| Tracking | High | High |
| PI tuning | Difficult | Less difficult |
| Complexcity | Medieum | High |
| THD | 1.19% | 0.41% |

## ACKNOWLEDGMENT

The Algerian Ministry of Higher Education and Scientific Research via the DGRSDT supported this research (PRFU Project code: A01L07UN190120180005).